\documentstyle[12pt]{article}

\makeatletter                    
\@addtoreset{equation}{section}  
\makeatother                     

\begin{document}

\title{Nonequilibrium Quantum Fields in Cosmology:
Comments on Selected Current Topics}
\author{B. L. Hu\thanks{ Email: hu@umdhep.umd.edu, hu@sns.ias.edu}\\
{\small Department of Physics, University of Maryland,
College Park, MD 20742, USA}\\
{\small Institute for Advanced Study, Princeton, New Jersey 08540, USA} }
\date{September 24, 1994}
\maketitle
\centerline{ (UMDPP-95-051,IASSNS-HEP-94/78) }
\centerline{\it Invited Talk given at the Second Journee Cosmologie,
Observatorie de Paris, June 2-4, 1994}

\begin{abstract}
Concepts of quantum open systems and ideas of correlation dynamics
in nonequilibrium statistical mechanics, as well as
methods of closed-time-path  effective action and influence functional
in quantum field theory
can be usefully applied for the analysis of quantum statistical
processes in gravitation and cosmology. We
raise a few conceptual questions and  suggest
some new directions of research on selected currrent topics on
the physics of the early universe, such as
entropy generation in cosmological particle creation,
quantum theory of galaxy formation, and phase transition
in inflationary cosmology.

\end{abstract}

\newpage

\section{Introduction}

The theme of this talk is to show how nonequilibrium statistical
field theory is useful, and in some cases, essential to analyzing
many problems in gravitation and cosmology, especially in black holes and
the early universe, from the Planck to the Grand Unified epochs.
There are a number of reviews I have
written in the last two years on this general subject matter, to which the
interested reader may refer. Each review  discusses some general issues and
then focuses on a specific topic. They are:\\

\noindent 1. Waseda Conference \cite{HuWaseda}: on quantum statistical
processes in the early
universe, with emphasis on noise in quantum fields and gravitational entropy.\\
2. Belgium \cite{HuBelgium}: on the quantum origin of noise and fluctuations
and the relevance of decoherence;  presents an influence functional approach
to the quantum theory of fluctuations and galaxy formation. For more recent
work, see \cite{nfsg,CHM}\\
3. Amsterdam  \cite{jr}: critical dynamics in the early universe:
ideas of scaling applied to the study of inflation; use of the quasilocal
effective potential for the study of slow-roll inflation.\\
4. Los Alamos \cite{HMLA}: the nature and the roles of quantum noise in
semiclassical gravity
and cosmology; particle creation as parametric amplification of quantum noise;
Hawking and Unruh radiation \cite{Haw,Unr}
derived from a statistical field theory viewpoint.
For details, see \cite{HM2,HM3}.\\
5. Banff \cite{HuBanff}: Nonequilibrium quantum field theory in gravitation;
examples with interacting fields and semiclassical cosmology. For details,
see \cite{fdrc,eft}\\

Let me mention some new developments in the last several months in these
directions. I'll  comment on the nature and focuses of the problems without
going into details, which can be found in the  quoted papers.

\section{ Particle Creation: Quantum Noise and Thermal Radiance}

Anglin \cite{Ang}, Hu and Matacz \cite{HM2},  Calzetta and Hu \cite{nfsg}
and Hu and Sinha \cite{fdrc} recently showed how
one can calculate particle creation in curved spacetimes via the
influence functional method \cite{if}.
In particlular, an expression for the influence
functional in terms of the Bogolubov coefficients was derived.
To place the old problem of particle creation in this new light enables one to
address statistical mechanical issues related to these quantum processes,
such as fluctuations and noise, decoherence, dissipation, backreaction,
and entropy generation in a unified framework \cite{HuTsukuba}.
Currently, Anglin, Hu and Raval \cite{AHR}
are analyzing the two detector problem, trying to resolve the controversy
raised by Grove, Raine and Sciama \cite{GRS} on whether an
inertial detector will detect radiation from an accelerated particle.
They concentrate on the field correlation as modified  by a moving
particle. The influence functional approach affords one the freedom
to examine the behavior of the particles or the field by coarse-graining
the information carried by the other (particle or field). Since it
includes the mutual effects  of the particles on and by the field,
the equations of motion of the subsystems (in the form of coupled Langevin
equations) contain the backreaction self-consistently.
New concepts are introduced, such as self- and mutual- noise and dissipation,
to describe the effective interaction of particles and field.

The conventional explanation of Unruh and Hawking radiation is that they
are special consequences of the existence of event horizons in uniformly
accelerating detectors  or black holes.
{}From physical considerations, cases with small departures from the uniform
should also emit radiations, albeit with a non-Planckian spectrum.
It is difficult to treat such cases from the  geometric viewpoint in terms of
event horizons. But
it should be natural from the statistical field theory viewpoint, as I argued
earlier \cite{HuEdmonton,HuPhysica} (my emphasis is thus closer in spirit
to the views of Bekenstein \cite{Bek} and Unruh \cite{UnrDH} ).
The influence functional formalism can treat any arbitrary state of motion.
There the radiation is seen to originate from
vacuum fluctuations excited by the motion of the particle, the limiting case
of uniform acceleration yielding an exact thermal radiance.
In statistical mechanics description, varying kinematical state of the detector
can alter in varying degrees the appearance of vacuum fluctuations
in Minkowsky space as a mixture of quantum and thermal noise in a different
space (only in the exact uniform acceleration case would it be describable
by the Rindler space). Currently Raval, Koks and Matacz \cite{HRKM}
are working out some examples  to illustrate these ideas I suggested
earlier \cite{HuEdmonton,HuPhysica,cgea}. This should lead to a unified
treatment and understanding
of thermal radiance from cosmological and black hole spacetimes.\\

%

\section{Entropy Generation in Cosmological Particle Creation}

In 1984, in discussing the conceptual problems of entropy generation
from particle creation I pointed out  \cite{Hu84} that the usual
simplistic identification of entropy with the number of created particles is
valid
only in the hydrodynamic (strongly interacting) regime. Theoretically,
for a free field, particle pairs created from the vacuum will remain in a pure
state and there should be no entropy generation. In 1986
Pavon and I \cite{HuPav} suggested that an intrinsic entropy of a
(free) quantum field can be measured
by the particle number (in a Fock space representation) or by the variance
(in the coherent state representation). It was pointed out that
the entropy of a (free) quantum field can change only if some information
of the field is lost or excluded from consideration, either by choosing
some special  initial state or by
introducing some measure of coarse-graining. For example,
the predicted monotonic increase in the spontaneous creation of
bosons is a consequence of the random phase initial condition implicitly
assumed in most discussions of vacuum particle creation.
\footnote{Parker \cite{Par69} first discussed the difference of spontaneous
and stimulated creation of bosons versus fermions.
Entropy generation in particle creation with interactions was discussed
by Hu and Kandrup \cite{HuKan}, the relation of random phase and particle
creation was further elaborated by Kandrup \cite{Kan88}.} For a recent
discussion of spontaneous and stimulated cosmological particle creation in
terms of squeezed states, see e.g., \cite{HKM}.

\subsection{The entropy of quantum fields}

Following these early discussions of the theoretical meaning of the entropy of
quantum fields, a recent upsurge of interest on this issue  was stimulated
by the work of Brandenberger, Mukhanov and Prokopec (BMP) \cite{BMP},
Gasperini and Giovannini (GG) \cite{GasGio} and others
on the entropy content of primordial graviton and other particles created.
The language of squeezed states for the description of cosmological
particle creation was introduced by Grishchuk and Sidorov \cite{GriSid}.
Though the physics is the same \cite{AFJP,HKM} as
was originally described by Parker \cite{Par69} and Zel'dovich \cite{Zel70}
the language makes it easier to compare with similar problems in quantum
optics,
which has delved into many similar theoretical and practical issues.
BMP suggested a coarse-graining  of the field by integrating out the rotation
angles, while GG used the dispersion in the superfluctuant operator.
In the large
squeezing limit (late times) they all get the same answer  $2r$ for the entropy
of spontaneously created particles, where $r$ is the squeezing parameter.
Here, as in the original work of Hu and Pavon, the choice of representation
of the state of the quantum field and the coarse-graining in the field are
stipulated, not derived.
The main point there is to show how the entropy of particle creation depends
on the choice of a specific initial state and/or a particular way of
coarse-graining.
How natural or physical this particular choice
is in a realistic cosmological context is not
addressed. This includes conditions when, for example, the quantum field is at
a finite temperature or is in disequilibrium, interacting with
other fields, or that its vacuum state is dictated by the choice of boundary
conditions in the earlier quantum cosmology regime (e.g., the Hartle-Hawking
boundary condition leading to the Bunch-Davies vacuum in de Sitter spacetime),
etc. To answer these questions, one needs to work with a larger theoretical
framework, that of statistical field theory of quantum open systems, as
I will now explain.

In the quantum Brownian motion (QBM) paradigm depiction of quantum field
theory in curved spacetime studied in the series of papers by Hu, Paz and Zhang
\cite{HPZ1,HPZ2}, and Hu and Matacz \cite{HM2},
the system represented by the Brownian particle can act as a detector
(as in the influence functional derivation of Unruh and Hawking radiation),
a particular mode of a quantum  field (such as the homogeneous inflaton
field), or the scale factor of the background spacetime (as in
minisuperspace quantum cosmology \cite{SinHu}), while the bath could be a set
of coupled
oscillators, a quantum field, or just the high frequency sector of the field,
as in stochastic inflation \cite{StoInf}.
The statistical properties of the system are
depicted by the reduced density matrix formed by integrating out the
details of the bath. Here one can use the reduced density matrix
or the associated Wigner function
to calculate the statistical averages of physical observables of the system,
and, in particular, the uncertainty or the entropy functions.
The time-dependence of the uncertainty function in a system interacting with
a heat bath illustrates the relative importance of
thermal and vacuum fluctuations and their roles in bringing about
the decoherence of the system and the emergence of classical behavior
\cite{HuZhaUncer,AndHal}. The entropy function for such open systems
was used \cite{ZHP,AndHal} as a measure of how close the description in terms
of different quantum states is to the classical dynamics.
For example, in a quantum Brownian model Zurek, Habib and Paz \cite{ZHP}
showed that the coherent state yields the minimal entropy, suggesting
the special relation it bears with the classical depiction.

\subsection{Entropy associated with the states of the system}

Here, the entropy function constructed from the reduced density matrix
(or the Wigner function) of a particular quantum state of the system
measures the information loss of the system in that state to the environment
(or, the `instability' characterized by the loss of predictive power relative
to the classical description  \cite{ZHP}).
One can study the entropy increase for a specific state, or compare
the entropy at each time for a variety of states characterized by the
squeeze parameter.
The uncertainty function plays a similar role \cite{HuZhaUncer,AndHal}. It
measures
the effects of  vacuum and thermal fluctuations in the environment
(at zero and finite temperature) on the observables of the system. The increase
of their variances due to these fluctuations gives rise to the uncertainty and
entropy increase.

Now what is the relation of this latter group of work to that of the former
and how the issues addressed here can help to answer the questions
I raised above?

The differences in appearance are obvious: The entropy of
\cite{HuPav,BMP,GasGio}
and others refers to that of the field, and is obtained by coarse-graining
some information of the field itself, such as making a random phase
approximation, adopting the number basis, or integrating over the rotation
angles. The entropy of the open system in  \cite{HuZhaUncer,AndHal,ZHP}
refers to that of the system and is obtained by coarse-graining the
environment.
Why is it that for certain generic models in some common limits (late time,
high squeeze), both groups of work obtain the same result?
For example, Matacz \cite{MatCohSta} considered a squeezed vacuum
of a time-dependent harmonic oscillator system, and motivated by the special
role of coherent states, modelled the effect of the environment by decohering
the squeezed vacuum in the coherent state representation.
He calculated the entropy function from this reduced density matrix,
and found that it approaches $2r$ in the high squeezing limit.
This calculation, though performed in the QBM open system framework, is in
spirit closer to the former work of field entropy in that the bath only
serves the token role for decohering the system and because the
time-dependent oscillator system admits particle creation in the normal modes
of the field.
In a way, approaching this problem from the open system viewpoint improves on
the
earlier work in that one can perform different coarse-grainings of the system
by changing its coupling to the environment.
What is more important conceptually, it clarifies the relation
between quantum and classical descriptions -- it is through decoherence
that the  field quanta behave like classical particles (loosely speaking).
\footnote{When one says he or she wants
to calculate the entropy content in classical matter or radiation observed
today
due to primordial particle creation, one needs to show the transition from
a quantum field to a classical matter description.
(This is similar to the questions on galaxy formation raised below,
specifically,
on the relation between quantum and classical fluctuations and how to get
classical
stochastic equations from quantum fields.)}
Thus it is not too surprising that the two groups of inquiries lead to similar
answers because the
same questions are asked about the oscillator (with time-dependent frequency)
as about the field.

Furthermore, an  even closer relation exists between the entropy of a system
and its environment theoretically. It can be shown that
any two subsystems of a closed system in a pure state will have equal entropy.
Indeed this is the reason why the derivation of black hole entropy can be
obtained equivalently by computing the entropy of the radiation
(e.g., \cite{FroNov}) emitted by the black hole, or by counting the
internal states (if one knows how!) of the black hole (e.g.,\cite{Bom}).
Here, if the open system and the environment together constitute
a closed system and they interact coherently, one expects that they would have
the same entropy \cite{Page,Sre}.
Physically one
can view what happens to the particle as a probe into the state of the field.
The application of open-system concepts to black hole entropy is a very
fruitful
avenue. (see the recent review of Bekenstein \cite{BekMG7})

\subsection{Directions}

Besides continuing the theoretical inquiries on uncertainty and entropy for
open
systems now pursued in earnest by Halliwell and coworkers \cite{Halentropy},
there are several directions one can advance on this issue of entropy
associated with cosmological particle production.
One currently undertaken by Koks and Matacz \cite{KMHR}
is to analyze the time-dependent harmonic oscillator interacting with
a bath of parametric oscillators.
One expects to find the entropy function to depend nonlocally on the entire
history of the squeezing parameter. This can be seen from the fact that
the rate of particle creation varies in time and its effect is history
dependent.
Existing methods of calculating the entropy
generation give results which only depend on the squeezing parameter at the
time of coarse-graining. There, both the choice of coarse-graining
and the time of its application are rather ad hoc and they affect the results
obtained.
The new method we are proposing has a more rigorous theoretical basis and
should provide more accurate estimates of entropy generation
from cosmological particle creation.

The other direction is to consider a bath of parametric oscillators
mimicking the field which produces particles,
and let the system be a particle detector probing the field. This picture
shifts the focus back on the field entropy again.
In the recent work of Calzetta, Hu and Matacz \cite{cgea,HM2,HMLA}
a physical link between quantum noise and particle creation is established,
and an expression for the influence functional in terms of the Bogolubov
coefficients is derived.
This formalism not only allows one to get everything familiar,
such as particle creation and interaction depicted in the established theories
(quantum fields in curved spacetimes)
but also new statistical mechanics information from first principles, such as
noise, as measured by the fluctuations in particle number, and entropy,
as constructed from the reduced density matrix.
One can also use the open system method to discuss entropy generation from
particle creation in the reheating phase. I will comment on this
at the end of my talk.


\section{Galaxy Formation from Quantum Fluctuations}

In \cite{HuBelgium}, Hu, Paz and Zhang pointed out some basic
deficiencies in
the existing theoretical framework of structure formation from quantum
fluctuations. One concerns the origin of noise, and the other
concerns the use of a classical stochastic equation for the description of
the dynamics of quantum fields. In that paper, they showed how to
derive the characteristics of quantum noise and
a classical stochastic equation from quantum field theory.
They found that colored multiplicative noise arise naturally from
fluctuations of interacting quantum fields, and that the
justification of using a classical stochastic equation for
the long wave-length modes depends on how successfully they decohere in the
face of these noises. The recent work of Calzetta and Hu \cite{nfsg}
raises yet a third important issue, questioning the validity of the
conventional use of the classical correlation functions for quantum fields
in these problems. Indeed because of the many ad hoc assumptions made in the
conventional theory of structure formation involving quantum fluctuations,
it is useful to reexamine the soundness of the whole theoretical foundation.
Let me raise a few general questions in this regard.

\subsection{Problems with the Existing Framework}

In the classical gravitational instability theory of galaxy formation
based on Lifshitz's 1946 work, the wave equation for the linear perturbations
of the background spacetime is driven by the first variations of the classical
energy momentum tensor. The scalar part of the perturbation is related
to the density contrast. It is often assumed that the initial source is
a white noise.  In the quantum theory of galaxy formation
a quantum scalar field (inflaton) mediates inflation, and its quantum
fluctuations seed the galaxies. Thus a) the vacuum energy density of the
scalar field drives the background spacetime into inflationary
expansion, while b) the fluctuations of the inflaton field produce the
density contrasts.
In stochastic inflation the long wavelength modes are driven by the
short wavelength quantum fluctuations. The folklore is that the former
behaves classically and the latter gives rise to a white noise. (We have
remarked that neither of these assumptions have been justified satisfactorily
\cite{cgea,HuBelgium}.)

Note first that level a) above invokes the semiclassical Einstein equation,
where the source is given by the vacuum expectation value of the energy
momentum tensor. In this case, it is the vacuum energy density of
the symmetric state before the system undergoes phase transition.
This is fine. At level b) there is confusion. Density contrast obtained
 from the first variation of the 00 component of the
energy momentum tensor is a classical object, but fluctuations of the scalar
field is quantum in nature.
In relating the two, one has presumably made a tacit assumption, i.e.,
that it is the average value of the quantum fluctuations, or something which
transforms them into classical fluctuations, that drives
the scalar part of the gravitational perturbation related to the
density contrast. What exactly turns the quantum fluctuations into
classical fluctuations is never made clear. Indeed, most authors seem not to be
bothered by this and just assume that the quantum fluctuations behave
like a classical stochastic source. We feel rather uncomfortable about this
state of affairs. In fact, in our investigations of noise and fluctuations
in semiclassical gravity, Calzetta and I \cite{nfsg} pointed out that this
is in general not the case.
Luckily there is a way to make these concepts and procedures
sound and rigorous via statistical field theory.

\subsection{New Theoretical Framework and Potential Problems}

The recent work of Calzetta, Hu, Matacz and Sinha \cite{nfsg,HM2,HM3,fdrc}
showed that the
semiclassical Einstein equation is only a mean-field theory. There is
actually a stochastic source term arising from the quantum fluctuations
of the matter field. It registers on the influence functional as the noise
kernel, which balances the dissipation kernel responsible for the backreaction
in the effective dynamics of the background geometry.
The semiclassical equation
which one customarily uses is the result of averaging over the noise
distribution. It is by means of this stochastic source, and within
this context, that the theory of galaxy formation based on
quantum fluctuations of matter fields can be made precise.
However, this new framework also raises a set of new issues.
For example, it was shown that quantum
fluctuations of the matter field in a dynamical spacetime are related to
particle creation.  When the excitation is below the
particle creation threshhold there is no stochastic source, but when it
is above the threshhold the stochastic source is related to
the difference in the amount of particle creation in neighboring histories.
Unlike classical fluctuations they don't come at all energy scales and
span the full spectrum. Worse yet, for conformal fields in conformally static
spacetimes,
like photons in a Robertson-Walker universe, there is no particle creation
and there should not be any stochastic source due to quantum fluctuations.
(Our preliminary analysis indicates that vacuum polarization terms like
the trace anomaly do not contribute to the stochastic source.)
For nonconformal fields, there is particle creation. But  during the
slow-roll inflationary epoch particle creation from the slowly changing
inflaton
field is small and concentrates in the low frequency modes.
(By contrast, in the reheating regime, the rapidly changing
field induces abundant particle creation and entropy generation, but
the spectrum will no longer be scale-invariant and there are other
problems.) The low production and weak fluctuations may not be
so bad in comparison with the observed inhomogeneity,
but the existing theoretical basis could be at peril if this were
true (we are cautiously guarded from making such a claim yet).
Further investigations on these issues are in place.

\subsection{Modelling and Analysis}

Going from the basic theoretical issues to modelling, recall that
in \cite{HuBelgium} a model of two interacting fields was used for
illustration.
One can use more realistic models for a thermal bath (as is in the
follow-up work of these authors \cite{qsf} in attempting to bridge their
stochastic field theory with the established thermal field theory),
or try to do the high-low frequency mode split
(as originally intended in the stochastic inflation program \cite{StoInf})
to derive a set of stochastic equations depicting more realistic conditions.
As for analysis, the functional Langevin equation deduced by us
can be simplified to an ordinary equation with some sampling functions
dictated by physical conditions. One can then be compare it with the
stochastic equations used for structure formation studies, and perform
numerical analysis.

This program of study while filling a gap
in the quantum theory of galaxy formation also can lead to new elements of
discovery, such as the effect of colored noise, non-Gaussian galaxy
distributions,
and anomalous correlations. It is interesting to compare the predictions
from different theoretical models with the limits drawn from the COBE data.

\section{Phase Transitions in the Early Universe}

There are at least three aspects in this subject:
a) the field theory aspect, which involves an infrared analysis
of the  effective action; b) the spacetime aspect, where the effects of
spacetime curvature and topology enter; and c) the statistical mechanical
aspect, dealing with nonequilibrium quantum fields.
(Only when they are in equilibrium can a finite temperature field theory be
useful). The field-theory aspect is at the heart of the matter. Many techniques
have been developed for treating infrared behavior, none could claim
perfection (or even accuracy). The approach of O'Connor and Stephens
who have made significant advances in the treatment of cross-over behaviors,
seems to me to be the most hopeful \cite{OCSrg}.
The spacetime aspect was studied with techniques developed in
quantum field theory in curved spacetime. The first stage of work between
1980-86 was reviewed in my  talk at the Fourth Marcel Grossmann
Meeting \cite{MG4}. In the work of Hu and O'Connor, and O'Connor et al
\cite{HuOC86,OSH} the Coleman-Jackiw-Tomboulis method \cite{CJT}
was used to treat the infrared behavior of quantum fields under
rather general conditions. Effects of curvature and topology on phase
transitions
can be understood in terms of finite size effect. Their
examples for symmetry breaking in product spacetimes contain many
interesting subcases, such as the imaginary-time finite temperature theory,
the Kaluza-Klein theory and Robertson-Walker universe.
The idea proposed by Hu and Zhang to understand inflation as scaling
\cite{cgea,jr} can provide some deeper insight into the physics of inflation
and black hole collapse. In the statistical mechanics aspect,
the 1988 work of Calzetta and Hu \cite{CH88} established a theoretical
foundation for studying non-equilibrium quantum fields.
The methods they proposed: the closed-time-path (CTP, or Schwinger-Keldysh)
functional formalism \cite{ctp},
the n-particle irreducible effective action, and the Wigner function
techniques,  have since been used as essential ingredients for many
statistical and kinetic field theory investigations into non-equilibrium
quantum processes, including particle creation, heavy-ion collision,
and quantum transport \cite{neqf}.
A quantum field theory of spinodal decomposition
was first studied by Calzetta \cite{CalSD} using the CTP formalism.
Recently Gleiser, Boyanovsky and others \cite{BoyGle} have applied these
concepts and techniques to study  the nature of the electroweak
phase transition (e.g., weakly first order), spinodal decomposition,
tunneling and domain wall formation with interesting results.
I refer to their talks in this conference.
Another recent development is the use of influence functional formalism to
describe noise and fluctuations. The recent work of Calzetta, Hu, Matacz, Paz,
Sinha and Zhang mentioned above
in applying the quantum Brownian model to field theory
provides a theoretical framework to study the effects of quantum fluctuations
and thermal noise on phase transition. I will find another occasion to
discuss this aspect. Here, instead, I would like to describe some thoughts
generated in my current work with Calzetta \cite{CH95} on
correlations and fluctuations 
in quantum field theory.

\subsection{Correlation and Noise in Kinetic Theory}

The relevant domain in the analysis of a system's approach to
a critical point is the infrared behavior of its fluctuations, and the
important object which carries this information is the correlation function.
A critical point is reached when the range of correlation becomes very large
(approaching infinity in bulk samples, while limited in finite size systems).
This was the starting point for our
earlier investigations into the symmetry behavior of quantum fields
in curved spacetimes (the work by Hu, O'Connor, Shen is summarized
in \cite{MG4}). In our current program of investigation \cite{CH95},
the objective is to extract the statistical information
of interacting quantum fields.
The infrared behavior of quantum fluctuations can be applied to phase
transition studies, but the formalism itself is useful for a broader range
of problems on relativistic kinetic processes.

The statistical mechanical paradigm used in our formalism is that of
Boltzmann's kinetic theory in the BBGKY hierarchy form. The key object is
the n-point correlation function.
This is different from the Brownian motion paradigm in the open system
conceptual framework, which has recently been successfully applied to the
discussion of decoherence, noise and dissipation problems in foundational
aspects
of quantum mechanics and semiclassical gravity. The philosophical
difference between these two major paradigms of non-equilibrium statistical
mechanics and their respective appropriateness in the application to different
physical problems have been discussed in the Introduction of \cite{CalHuDCH}.
There we proposed the correlation history as a more natural way of
coarse-graining in the decoherent history formulation of quantum mechanics
\cite{dechis}.
One of the lessons we learned in adopting the open system way of thinking is
how to treat noise and fluctuations rigorously from first principles.
In our present work
we want to incoporate noise
and fluctuations into dissipation, all described in terms of the hierarchy of
correlation functions. Thus, concepts from both paradigms are utilized.

Many new results come from this line of inquiry. Let me just mention two,
one concerns effective field theory, and the other concerns stochastic
mechanics. The former addresses renormalization group theory
and gauge hierarchy problems, and the latter addresses the source and nature of
fluctuations in quantum fields, both are relevant to phase transition problems.

How does one define noise in the correlation framework? In an open system
one coarse-grains the environment and obtains noise. Here there is no clear
separation of the system and the environment, as all particles in the
kinetic theory are treated on equal footing. A separation is possible
by the correlation order. If one keeps all orders one has complete information,
just as the complete BBGKY hierarchy is equivalent to the full set of Newton's
equations for the n particles. In actual measurements, one does not have the
ability to keep track of all orders, and usually settles with the one-particle
distribution function and perhaps the two- or three- particle correlation
functions. The description of the total system is therefore incomplete by
choice. Recall that the BBGKY hierarchy has the dynamics of an n-th order
correlation function driven by a source of n+1 th order functions, ad
infinitum.
Only when one assumes that the n+1 th order correlation function factorizes
(into products of lower order correlation functions) will there be loss of
information and the appearance of dissipation in the effective dynamics
of the nth order correlation function.
Physically, this is the molecular chaos assumption which
Boltzmann used to explain dissipation in Hamiltonian dynamics and to
postulate the H-theorem. One can define noise of the nth correlation order
as the source term (formed by the n+1 th order correlation functions) in the
nth
order BBGKY equation. Usually this term is ignored when the hierarchy is
simply truncated-- like the collisionless Boltzmann equation. But the
correlation noise has a place to itself, both in practical and theoretical
terms.
Practically, this is the stochastic source for the lower order equations,
and theoretically, it is needed in the balance of information, or loss thereof.
These two points are easier
to see in the open system framework, the former via the Langevin equation and
the latter via the fluctuation-dissipation relation. But they
are also there in the kinetic theory framework.

\subsection{Fluctuations and Correlation of Quantum Fields}

In field theory language, the full set of nth order correlation functions
enters
into the n-particle irreducible (nPI) effective action. Complete
information for the system is contained in the $\infty PI$, or what we call
the master effective action.
At any order determined by the level of accuracy of the experiment,
the effective equation for the correlation function is driven by a noise term
which contains the information of all the higher order correlations.
Truncation at any level by factorization discards this information and
introduces  dissipation in the effective dynamics of that order.
The ordinary effective action we encounter in field theory is a functional of
the mean field. There is no room for the noise terms associated with
quantum fluctuations. For phase transition studies one usually includes
the 2-particle correlation function (there is of course no guarantee that
this is sufficient to depict the infrared behavior-- usually they don't --
witness the frustration in the higher order calculations of the electroweak
phase transition.) The relation between the mean field and the correlation
function(s) was not clearly treated in the old way. But cast in this new
statistical field theoretical framework, one can actually calculate the
stochastic source term driving the correlation functions at each level, and
verify their consistency by the fluctuation-dissipation relations at
each level. In this light one can also understand why effective theories
can be intrinsically dissipative in nature \cite{HuPhysica,HuBanff,eft}. These
stochastic equations can provide a new basis for the analysis of nucleation
and spinodal decomposition processes, as the dynamics of fluctuations
and correlations can be studied on the same footing.
The implications of this new statistical field theory on renormalization group
theory and phase transitions are under investigation \cite{eft,CH95,CHP}.

\subsection{Dynamics of the Inflaton Field}

The dynamics of inflationary cosmology depends sensitively on how the
order parameter field (inflaton) evolves, from the symmetric to the
broken symmetry phase, i.e., to enter a vacuum-energy dominated phase
(e.g., needs a metastable potential), to sustain the inflation (e.g.,
slow-roll)
and to gracefully exit (reheating). The usual treatment is via a
finite-temperature effective potential which a priori assumes an
equilibrium condition. But the time-dependence of the inflaton in a dynamic
spacetime really calls for a fully non-equilibrium treatment.
As we pointed out before, this is
one of the three major factors lacking in the usual treatment
of phase transitions in the early universe \cite{MG4} (i.e., field theory
infrared behavior, geometric and topological effects,  and statistical
mechanical effects).
With a stochastic field theory set up we can now improve upon this aspect
of the problem. For example, in the study of inflaton dynamics
by Guth and Pi \cite{GutPi} using a quantum mechanical inverted
harmonic oscillator model, the transition to the classical regime was defined
via a simple uncertainty principle argument. One can actually calculate
the transition time by comparing the relative magnitude of the quantum and
thermal fluctuations based on our recent result on decoherence \cite{HPZ1} and
the uncertainty principle \cite{HuZhaUncer}. This was done by Raval recently
\cite{HuRav}. Cornwall and Bruinsma \cite{CorBru} first suggested
using the influence functional method for this problem but oversimplified
the problem in order to cater to the  Calderia-Leggett model.
One can provide a more realistic and in-depth depiction of the inflaton
dynamics
with the statistical field theory results \cite{qsf}.

\subsection{Reheating}

The standard picture of reheating is given qualitatively by Dolgov and
Linde \cite{DolLin}, Abbott, Farhi and Wise \cite{AFW}.
There is a recent revival of interest in the reheating phase of inflation
stimulated by the work of Kofman, Linde and Starobinsky \cite{KLS} and
Shtanov, Traschen and Brandenberger \cite{STB}. I just want to
call the attention to the fact that there is a rigorous way of calculating
particle creation and interaction in the
phase where the inflaton field undergoes rapid oscillation and damping.
The method is the Schwinger-Keldysh (or closed-time-path) \cite{ctp}
effective action formalism, which has been applied to cosmological particle
creation and backreaction problems before \cite{CalHu87,CalHu89}.
Specifically for the reheating problem, Paz \cite{Paz90} and Stylianopoulis
\cite{StyPhD} have used this method to study particle creation and decay
in field theory models with quadratic and Yukawa type couplings.
Their calculation also included the backreaction of created particles which
is encapsuled in the viscosity function in the effective equation of motion
for the inflaton field. Making use of the intimate relation of the CTP
effective action with the influence functional formalisms \cite{cgea,fdrc}
one can also deduce the noise kernel
associated with particle creation in the reheating phase and a fluctuation
dissipation relation for such processes. This is a far more
sophisticated and rigorous approach to this problem than the time-dependent
perturbation theory used in the older works.
With it one can address the issues
raised in \cite{KLS,STB} and discern the physical scenarios based on
quantitative rather than descriptive results. The recent work of Boyanovsky,
de Vega, Holman, Lee and Singh \cite{BDHLS} seems to me to be the most advanced
in the analysis of this problem. \\

\noindent{\bf Acknowledgement}
I thank Esteban Calzetta and Andrew Matacz for discussions on various topics
touched upon in this talk.
The kind hospitality of Professors de Vega and Sanchez made my short stay
in Paris a very pleasant one.
This work is supported in part by the National Science Foundation under
grant PHYS91-19726, the General Research Board of the Graduate School of
the University of Maryland and the Dyson Visiting Professor Fund at the
Institute for Advanced Study, Princeton.


\begin{thebibliography}{999}

\bibitem {HuWaseda}
B. L. Hu, ``Quantum Statistical Processes in the Early Universe"
in {\it Quantum Physics and the Universe}, Proc. Waseda Conference, Aug. 1992
ed. M. Namiki et al  (Pergamon Press, Tokyo, 1993).
Vistas in Astronomy 37, 391 (1993)

\bibitem {HuBelgium}
B. L. Hu, J. P. Paz and Y. Zhang ``Quantum Origin of Noise and
Fluctuations in Cosmology'',
in {\it The Origin of Structure in the Universe}, edited by E. Gunzig and
P. Nardone (Kluwer, Dordrecht, 1993), p. 227.

\bibitem {nfsg}
E. Calzetta and B. L. Hu,``Noise and Fluctuations in Semiclassical Gravity",
Phys. Rev. D49, 6636 (1994)

\bibitem {CHM}
E. Calzetta, B. L. Hu and A. Matacz, in preparation

\bibitem{jr}
B. L. Hu, Class. Quan. Grav. 10, S93 (1993)

\bibitem  {HMLA}
B. L. Hu and A. Matacz, ``Quantum Noise in Gravitation and Cosmology"
Invited Talk at the Workshop on {\it Fluctuations and Order}, ed. M. Millonas
(Springer Verlag, Berlin, 1994). Univ. Maryland preprint pp94-44 (1993)
 astro-ph/9312012

\bibitem{Haw}
S. W. Hawking, Comm. Math. Phys. 43, 199 (1975)

\bibitem{Unr}
W. G. Unruh, Phys. Rev. D14, 870 (1976)

\bibitem {HM2}
B. L. Hu and A. Matacz, ``Quantum Brownian Motion in a Bath of Parametric
Oscillators", Phys. Rev. D49, 6612 (1994)

\bibitem {HM3}
B. L. Hu and A. Matacz, ``Backreaction in
Semiclassical Cosmology: the Einstein-Langevin Equation",
Univ. Maryland preprint 94-31 (1993). gr-qc/9403043

\bibitem{HuBanff}
B. L. Hu, ``Quantum Statistical Fields in Gravitation and Cosmology"
in Proc. Third International Workshop on Thermal Field Theory and
Applications, eds. R. Kobes and G. Kunstatter
(World Scientific, Singapore, 1994) gr-qc/9403061

\bibitem {fdrc}
B. L. Hu and S. Sinha, ``Fluctuation-Dissipation Relation in Cosmology",
   Univ. Maryland preprint pp93-164 (1993) gr-qc/9403054

\bibitem{eft}
E. Calzetta, B. L. Hu and Yuhong Zhang,
``Dissipative Nature of Effective Field Theories" (in preparation)

\bibitem {Ang}
J. R. Anglin, Phys. Rev. D47, 4525 (1994)

\bibitem{if}
R. Feynman and F. Vernon, Ann. Phys. (NY) {\bf 24}, 118 (1963).
R. Feynman and A. Hibbs, {\it Quantum Mechanics and Path Integrals},
(McGraw - Hill, New York, 1965).
A. O. Caldeira and A. J. Leggett, Physica {\bf 121A}, 587 (1983);
Ann. Phys. (NY) {\bf 149}, 374 (1983).
H. Grabert, P. Schramm and G. L. Ingold, Phys. Rep. {\bf 168}, 115 (1988).
B. L. Hu, J. P. Paz and Y. Zhang, Phys. Rev. {\bf D45}, 2843 (1992);
{\bf D47}, 1576 (1993)

\bibitem{HuTsukuba}
B. L. Hu, in {\it Thermal Field Theories}, eds. H. Ezawa et al (North-Holland,
Amsterdam, 1991) p.223

\bibitem {AHR}
A. Raval, B. L. Hu and J. R. Anglin , in preparation

\bibitem {GRS}
P. G. Grove, Class. Quan. Grav. {\bf 3}, 801 (1986)
D. J. Raine, D. W. Sciama, and P. G. Grove, Proc. Roy. Soc. Lond.
 {\bf A435}, 205 (1991)

\bibitem {HuEdmonton}
B. L. Hu, in {\it Proc. CAP-NSERC 1987 Summer Institute in Theoretical Physics}
eds G. Kunstatter et al, (World Scientific, Singapore, 1988) Vol 2, p. 252-276

\bibitem {HuPhysica}
B. L. Hu, Physica A158, 399 (1989).

\bibitem {cgea}
B. L. Hu and Y. Zhang, ``Coarse-Graining, Scaling, and Inflation"
Univ. Maryland Preprint 90-186 (1990);
B. L. Hu, in {\it Relativity and Gravitation: Classical
and Quantum} Proc. SILARG VII, Cocoyoc, Mexico 1990.
eds. J. C. D' Olivo et al (World Scientific, Singapore 1991).

\bibitem {Bek}
J. D. Bekenstein, Phys. Rev. D12, 3077 (1975).

\bibitem {UnrDH}
W. G. Unruh, Phys. Rev. Lett. 46, 1351 (1981);
``Dumb Holes", gr-qc/9409008 (1994)


\bibitem {HRKM}
B. L. Hu, A. Raval, D. Koks, A. Matacz, in preparation

\bibitem{Hu84}
B. L. Hu, in {\it Cosmology of the Early Universe} eds. L. Z. Fang and R.
Ruffini
(World Scientific, Singapore, 1984)

\bibitem {HuPav}
B. L. Hu and D. Pavon, Phys. Lett. {\bf B180}, 329 (1986)

\bibitem{Par69}
L. Parker, Ph. D. Thesis, Harvard University, 1966; Phys. Rev. Lett. 21, 562
(1968); { Phys. Rev. } {\bf 183}, 1057 (1969); {\bf D3}, 346 (1971)

\bibitem{Zel70}
Ya. B. Zel'dovich, Pis'ma Zh. Eksp. Teor. Fiz, {\bf 12} ,443 (1970)
[JETP Lett. {\bf 12}, 307(1970)];

\bibitem {HuKan}
B. L. Hu and H. E. Kandrup, Phys. Rev. {\bf D35}, 1776 (1987)

\bibitem {Kan88}
H. E. Kandrup, Phys. Rev. {\bf D37}, 3505 (1988)

\bibitem {HKM}
B. L. Hu, G. W. Kang and A. Matacz, Int. J. Mod. Phys. A9, 991 (1994)

\bibitem {BMP}
R. H. Brandenberger, V. Mukhanov and T. Prokopec,
Phys. Rev. Lett. 69, 3606 (1992); Phys. Rev. D48, 2443 (1993)

\bibitem {GasGio}
M. Gasperini and M. Giovanni, Phys. Lett. 301B, 334 (1993);
M. Gasperini and M. Giovanni and Veneziano, Phys. Rev. D48, R439 (1993).

\bibitem {GriSid}
L. Grishchuk and Y. V. Sidorov, Phys. Rev. D42, 3414 (1990)

\bibitem {AFJP}
A. Albrecht, P. Ferreira, M. Joyce and T. Prokopec, Imperial College
preprint TP/92-93/21, astro-ph/9303001


\bibitem {HPZ1}
B. L. Hu, J. P. Paz and Y. Zhang, Phys. Rev. {\bf D45}, 2843 (1992)

\bibitem {HPZ2}
B. L. Hu, J. P. Paz and Y. Zhang, Phys. Rev. {\bf D47}, 1576 (1993)

\bibitem {SinHu}
Sukanya Sinha and B. L. Hu, Phys. Rev. D44, 1028 (1991)
\bibitem{StoInf}
A. A. Starobinsky, in {\it Field Theory, Quantum Gravity and Strings},
ed. H. J. de Vega and N. Sanchez (Springer, Berlin 1986);
J. M. Bardeen and G. J. Bublik, Class. Quan. Grav. {\bf 4}, 473 (1987).


\bibitem {HuZhaUncer}
B. L. Hu and Yuhong Zhang, Mod. Phys. Lett. A8, 3575 (1993);
Univ. Maryland preprint 93-162 (1993), gr-qc/93012034
B. L. Hu and Yuhong Zhang, in Proc. Third International
Workshop on Quantum Nonintegrability, Drexel University, Philadelphia,
May 1992, eds J. M. Yuan,  D. H. Feng and G. M. Zaslavsky
(Gordon and Breach, Langhorne, 1993).

\bibitem {AndHal}
A. Anderson and J. J. Halliwell, Phys. Rev. {\bf D48}, 2753 (1993).

\bibitem {ZHP}
W. H. Zurek, S. Habib and J. P. Paz, Phys. Rev. Lett. 70, 1187 (1993)

\bibitem {MatCohSta}
A. Matacz, Phys. Rev. D49, 788 (1994)



\bibitem {FroNov}
V. Frolov and I. D. Novikov, Phys. Rev. D48, 4545 (1993)


\bibitem {Bom}
L. Bombelli, R. K. Koul, J. Lee and R. D. Sorkin, Phys. Rev. D34, 373 (1986)

\bibitem {Page}
D. N. Page, Phys. Rev. Lett. 71, 1291 (1993)

\bibitem {Sre}
M. Srednicki, Phys. Rev. Lett. 71, 666 (1993)

\bibitem {BekMG7}
J. D. Bekenstein, Review Talk at MG7 (1994) gr-qc/9409015


\bibitem {Halentropy}
J. J.  Halliwell, Phys. Rev. D48, 2739, 4785 (1993);
C. Anastopoulos and J. J. Halliwell, Imperial College Preprint
IC 93-94/53 (1994) gr-qc/9407039

\bibitem {KMHR}
D. Koks, A. Matacz, B. L. Hu and A. Raval, in preparation

\bibitem {qsf}
B. L. Hu, J. P. Paz and Y. Zhang, ``Stochastic Dynamics of
Interacting Quantum Fields'' in preparation (1994)

\bibitem{OCSrg}
Denjoe O'Connor and C. R. Stephens, ``Environment-Friendly Renormalization
Group Theory" Int. J. Mod. Phys. (1994)

\bibitem{MG4}
B. L. Hu,``Phase Transition in the Early Unvierse: Geometric Effects" in
{\it Recent Developments in General Relativity: Proc. 4th Marcel Grossmann
Meeting, Rome, 1985} ed. R. Ruffini (North Holland, Amsterdam, 1986)

\bibitem{HuOC86}
B. L. Hu and D. J. O'Connor, Phys. Rev. {\bf D36}, 1701 (1987).

\bibitem{OSH}
D. J. O'Connor, C. R. Stephens and B. L. Hu, Ann. Phys. (N.Y.)
{\bf 190}, 310 (1990).

\bibitem {CJT}
J. M. Cornwall, R. Jackiw and E. Tomboulis, Phys. Rev. D10, 2428 (1974)

\bibitem{ctp}
J. Schwinger, J. Math. Phys. {\bf 2} (1961) 407;
P. M. Bakshi and K. T. Mahanthappa, J. Math. Phys. 4, 1 (1963), 4, 12 (1963).
L. V. Keldysh, Zh. Eksp. Teor. Fiz. {\bf 47 }, 1515 (1964)
[Engl. trans. Sov. Phys. JEPT {\bf 20}, 1018 (1965)].
G. Zhou, Z. Su, B. Hao and L. Yu, Phys. Rep. {\bf 118}, 1 (1985);
Z. Su, L. Y. Chen, X. Yu and K. Chou, Phys. Rev. {\bf B37}, 9810 (1988).
B. S. DeWitt, in {\it Quantum Concepts in Space and Time},
ed. R. Penrose and C. J. Isham (Claredon Press, Oxford, 1986);
R. D. Jordan, Phys. Rev. D33, 44 (1986).
E. Calzetta and B. L. Hu, Phys. Rev. {\bf D35}, 495 (1987);
{\bf D37}, 2878 (1988); Phys. Rev. {\bf D40}, 656 (1989).

\bibitem{CH88}
E. Calzetta and B. L. Hu, Phys. Rev. {\bf D37}, 2878 (1988).

\bibitem{neqf}
See, e.g., F. Cooper, et al, hep-ph/9404357; Los Alamos Preprint
LA-UR-94 783 (1994)

\bibitem{CalSD}
E. Calzetta, Ann. Phys. (N.Y.) 190, 32 (1989)


\bibitem{BoyGle}
D. Boyanovsky and H. J. de Vega, Phys. Rev. D47, 2343 (1993);
D. Boyanovsky, D. S. Lee and  A. Singh, Phys. Rev. D48, 800 (1993);
D. Boyanovsky, H. J. de Vega, and R. Holman, Univ. Pittsberg preprint 93-6;
M. Gleiser, G. C. Marques and R. O. Ramos, Phys. Rev. D48, 1571 (1993);
M. Gleiser and R. O. Ramos, Univ. Dartmouth preprint DART-HEP-93/06 (1993)

\bibitem{Lef}
W. Horsthemke and R. Lefever, {\it Noise Induced Transitions} (Springer,
Berlin 1984)

\bibitem {CH95}
E. Calzetta and B. L. Hu, "Correlation, Noise and Fluctuations in
Interacting Quantum Fields" (1994)

\bibitem{CalHuDCH}
E. Calzetta and B. L. Hu, ``Decoherence of Correlation Histories''
in {\it Directions in General Relativity, Vol II: Brill Festschrift},
eds B. L. Hu and T. A. Jacobson (Cambridge University Press, Cambridge, 1993)

\bibitem {dechis}
R. Omn\'es, Rev. Mod. Phys. {\bf 64}, 339 (1992);
J. B. Hartle, ``Quantum Mechanics of Closed Systems"
in {\it Directions in General Relativity, Vol. 1: Misner Festschrift},
eds B. L. Hu, M. P. Ryan and C. V. Vishveswara
(Cambridge Univ., Cambridge, 1993).

\bibitem {CHP}
E. Calzetta, B. L. Hu and J. P. Paz, in preparation

\bibitem {GutPi}
A. H. Guth and S.-Y. Pi, Phys. Rev. D32, 1899 (1985)

\bibitem {HuRav}
B. L. Hu and A. Raval, in preparation

\bibitem {CorBru}
J. M. Cornwall and R. Bruinsma, Phys. Rev. {\bf D38}, 3146 (1988)


\bibitem {DolLin}
A. D. Dolgov and A. D. Linde, Phys. Lett. 116B, 329 (1982)

\bibitem {AFW}
L. Abbott, E. Farhi and M. Wise, Phys. Lett. 117B, 29 (1982)

\bibitem {KLS}
L. A. Kofman, A. D. Linde and A. A. Starobinsky, ``Reheating After Inflation"
hep-th/9405187 (1994)

\bibitem {STB}
Y. Shtanov, J. Traschen and R. Brandenberger, ``Universe Reheating After
Inflation"
Brown University preprint HET-957 (1994)


\bibitem {CalHu87}
E. Calzetta and B. L. Hu, Phys. Rev. {\bf D35}, 495 (1987).

\bibitem {CalHu89}
E. Calzetta and B. L. Hu, Phys. Rev. {\bf D40}, 656 (1989).

\bibitem{Paz90}
J. P. Paz, Phys. Rev. {\bf D42}, 529 (1990)

\bibitem {StyPhD}
A. Stylianopoulos, Ph. D. Thesis, University of Maryland (1991)

\bibitem {BDHLS}
D. Boyanovsky, H. J. de Vega, R. Holman, D.-S. Lee and A. Singh,
``Dissipation via Particle Production in Scalar Field Theories"
(1994) hep-ph/9408214

\end{thebibliography}
\end{document}